\documentclass[letterpaper, 10 pt, conference]{ieeeconf}

\IEEEoverridecommandlockouts                              

\usepackage{comment}
\usepackage{cite}
\usepackage{amsmath,amssymb,amsfonts}
\usepackage{algorithmic}
\usepackage{graphicx}
\usepackage[export]{adjustbox}
\usepackage{textcomp}

\title{\LARGE \bf A Dynamic Model and Stability Analysis of a Grid-Interactive Data Center}

\author{Advith Natarajan, Xinyang Zhou, Guido Cavraro
\thanks{A. Natarajan is wiith the Electrical Engineering and Computer Sciences department, UC Berkeley. Email: {\tt advith@berkeley.edu}}
\thanks{X. Zhou and G. Cavraro are with the Power Systems Engineering Center, National Laboratory of the Rockies. Email: {\tt xinyang.zhou@nlr.gov, guido.cavraro@nlr.gov}}
\thanks{This work was authored in part by the National Laboratory of the Rockies for the U.S. Department of Energy (DOE) under Contract No. DE-AC36-08GO28308. Funding provided by the Advanced Modeling Grid Research Program at the Office of Electricity, U.S. Department of Energy. The views expressed in the article do not necessarily represent the views of the DOE or the U.S. Government. The U.S. Government retains and the publisher, by accepting the article for publication, acknowledges that the U.S. Government retains a nonexclusive, paid-up, irrevocable, worldwide license to publish or reproduce the published form of this work, or allow others to do so, for U.S. Government purposes.}
}

\begin{document}
\maketitle

\begin{abstract}
The rapid growth of data centers is creating new interactions between computational workload and power-system dynamics, with potentially significant implications for grid stability. We develop a four-dimensional nonlinear dynamical model of a data center campus interconnected with the bulk power system through a modified Single Machine Infinite Bus  system. The model couples generator swing dynamics with a fluid-limit model of computational workload backlog and a grid-interactive data center power controller that incorporates both frequency feedback and computational backpressure. This formulation captures the coupled evolution of generator frequency, rotor angle, computational workload, and data center power consumption, providing a framework for studying the stability implications of flexible data center operation.
\end{abstract}

\section{Introduction}
The rapid deployment of large data centers (DCs) is transforming the interaction between computational infrastructure and the electric power system, introducing new sources of large and potentially highly variable demand that challenge grid planning, operation, and stability~\cite{kercci2024impact}. At the same time, DC deployment plays a critical role in maintaining U.S. leadership in AI, making the integration of DCs an increasingly important challenge for the power system.

Several research efforts have focused on developing DC power demand models, ranging from individual computing components such as CPUs and GPUs~\cite{Rajput} to aggregated facility-level models~\cite{ko2026,vercellino2026measurement}. 
Another research direction focuses on developing detailed digital twins of DCs for numerical simulation and stability analysis~\cite{Athavale}, including high-fidelity electromagnetic transient (EMT) models~\cite{Brett}. Nevertheless, such models typically do not explicitly connect computational workload dynamics with grid-level quantities, thereby missing a key characteristic that distinguishes DCs from conventional large loads, such as industrial facilities or other large commercial campuses.

A recent line of research is exploring how DCs operations affect the power system and how DC can become \emph{grid interactive}. DCs can leverage their inherent flexibility, including UPS battery systems, cooling systems, and deferrable computational workloads, to mitigate their impact on the grid~\cite{colangelo2026ai} or respond to energy prices~\cite{Xinyang_DC2026}. By dynamically adjusting their power consumption in response to grid conditions, DCs could potentially contribute to grid stability while maintaining the computational performance required by their workloads. For example, \cite{Gupta} derives frequency-security specifications for DC loads; \cite{Yan} proposes decentralized voltage control that exploits the phase-dependent structure of AI training workloads. The work~\cite{shenoy2026stabilitypowersystemsai} characterizes the grid stability under switching DC power demand resulting from training jobs. However, these approaches generally treat the DC power demand as an externally specified input or scheduling quantity, rather than as a dynamic state coupled to an evolving computational workload and the electromechanical dynamics of the power system.

In this paper, we develop a four-dimensional nonlinear dynamical model of a grid-interactive DC campus consisting of a DC facility and a local generating unit. The model is formulated by modifying a Single Machine Infinite Bus (SMIB) system, in which the bulk power system is represented by an infinite bus.
The electrical dynamics are coupled with a fluid-limit model describing the evolution of the DC's computational-work backlog. 
The DC power demand determines the rate at which computational work is processed and is regulated through a feedback control rule that balances computational requirements with the operating conditions of the surrounding power system. 
The proposed formulation therefore captures the dynamic interaction among grid frequency, DC power consumption, and computational workload.
The resulting nonlinear system can be used to analyze the stability of the grid under different power consumption protocols. We evaluate different DC control parameters and compare the certified regions of attraction under these different conditions. The results provide a framework for quantifying how DC flexibility and grid-aware control can affect the stability of the coupled power-system and computational dynamics.

\section{Grid-Interactive DC Modeling}
\label{sec:system_model}

\begin{figure}[tb]
    \centering
    \includegraphics[width=0.7\linewidth]{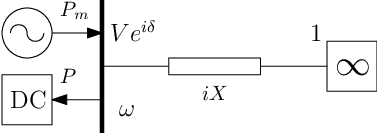}
    \caption{DC--SMIB system architecture.}
    \label{fig:datacenter_smib_diagram}
\end{figure}

We model the grid-interactive DC using the modified SMIB system shown in Figure~\ref{fig:datacenter_smib_diagram}. The DC campus, consisting of a local generating unit and the DC facility, is connected to a common electrical bus, which is interconnected with the transmission grid, represented by an infinite bus, through a purely inductive line with impedance $iX$. The infinite-bus voltage phasor is taken as the reference, with magnitude $1$ p.u., while the voltage phasor at the DC bus is represented as $V e^{\mathrm{i}\delta}$. We assume that the local generating unit regulates reactive power to maintain the voltage magnitude $V$ at the DC bus constant.

The resulting system is described by four state variables:
\begin{itemize}
    \item the generator rotor angle relative to the synchronous reference frame, $\delta$, measured in radians. For simplicity, we assume that the DC-bus voltage angle coincides with the generator rotor angle, so that the same variable $\delta$ describes both the generator angle relative to the synchronous reference and the phase angle of the DC-bus voltage phasor;
    \item the generator rotor angular frequency, $\omega$, measured in radians per second, with nominal synchronous speed $\omega_0=2\pi(60)\,\mathrm{rad/s}$;
    \item the active power consumed by the DC facility, $P$, expressed in Watts (W); 
    \item the computational-work backlog at the DC, $q$, measured in floating-point operations (FLOP).
\end{itemize}
These variables are collected in the state vector 
$x =\begin{bmatrix}
\delta, & \omega, & P, & q    
\end{bmatrix}^\top$. The state dynamics are:
\begin{subequations}
\begin{align}
& \dot\delta = \omega-\omega_0, \label{eq:delta}\\
& J\omega\dot\omega+D\omega(\omega-\omega_0)
=P_m-P-\gamma\sin\delta, \label{eq:swing}\\
& \tau \dot P = P_0 + f - P, \label{eq:pdc}\\
& \dot q = \lambda - \mu(P - P_0). \label{eq:q}
\end{align}
\label{eq:4d_mod}
\end{subequations}
Equations~\eqref{eq:delta}--\eqref{eq:swing} describe the electromechanical dynamics of the synchronous generator and its interaction with the DC load. Rather than using the conventional simplified swing equation~\cite{Cavraro_DC2026}, we adopt the improved swing equation proposed in~\cite{monshizadeh2016}, which retains the explicit dependence of the inertial and damping terms on the rotor speed $\omega$. The parameters $J$ and $D$ denote the generator rotor moment of inertia and damping coefficient, respectively, and $P_m$ is the mechanical power input. The parameter $\gamma:=V/X$ represents the maximum active-power transfer through the interconnecting reactance $X$, under the assumed voltage magnitudes.

Equation~\eqref{eq:pdc} describes the dynamics of the DC power demand. The time constant $\tau$ captures the aggregate response time associated with changes in power consumption. The parameter $P_0$ captures the baseline power demand associated with non-computational facility functions and non-flexible computational power demand, while $f$ represents the controllable computational power demand set point designed to make the DC operation responsive to grid conditions. Notice that a first-order model of DC power demand has also been proposed in~\cite{Milano2026}. Since we assume that $P_0$ is fixed, the instantaneous power allocated to computation is given by $P - P_0$.

Equation~\eqref{eq:q} models the computational-work dynamics and captures the coupling between electrical power consumption and computational throughput. Although individual computational tasks arrive and complete as discrete events, we adopt a fluid-limit approximation in which the computational workload is represented by the continuous state $q$. Under appropriate scaling assumptions, stochastic multi-server queueing systems can be approximated by deterministic fluid-limit dynamics~\cite{gromoll2002}. Related models have also been used to describe aggregate computational resources under load-balancing and auto-scaling regimes~\cite{mukherjee2017,yekkehkhany2020}. The parameter $\lambda$ denotes the rate at which computational work enters the system, while $\mu(P - P_0)$ denotes the computational-work service rate associated with the computational-power demand $P - P_0$. Both quantities are measured in FLOP/s.

\section{Grid-Aware Control and Equilibrium Analysis}

In the remainder of the paper, we specialize the general model~\eqref{eq:4d_mod} by specifying the functional forms of the computational service rate and the control input $f$. Although the underlying computing infrastructure consists of discrete hardware resources executing individual computational tasks, the aggregate relationship between computational power and throughput can be approximated as continuous over the operating range of interest. In particular, we adopt a linear power-to-throughput relationship consistent with the empirical characterization reported in~\cite{chen2026}:
\begin{equation}
\mu(P-P_0)=\kappa (P - P_0),
\label{eq:kP}
\end{equation}
where $\kappa$, measured in $\mathrm{FLOP}/(\mathrm{s}\cdot\mathrm{W})$, denotes the effective conversion factor from computational power to computational throughput. Here, $P-P_0$ represents the computational power consumed by the DC, excluding the baseline non-computational facility demand $P_0$.

We define the computational-power setpoint $f$ as a function of the generator frequency and computational-work backlog:
\begin{equation}
f(\omega,q) = \alpha q + \beta(\omega-\omega_0),
\label{eq:control}
\end{equation}
where $\alpha>0$ and $\beta\geq0$ are control gains.
The first term in~\eqref{eq:control} implements a computational backpressure mechanism. As the computational-work backlog $q$ increases, the controller raises the computational-power setpoint, thereby increasing the computational service rate once the actual power demand responds. This mechanism couples computational workload to electrical power consumption and helps prevent persistent workload accumulation when sufficient computational capacity is available. 
The second term implements a frequency-responsive droop mechanism. When the grid frequency decreases below its nominal value, the controller reduces the computational-power setpoint, thereby providing load relief during under-frequency conditions. Conversely, when the frequency rises above its nominal value, the controller increases the setpoint, allowing the DC to absorb additional power when generation exceeds demand.

\begin{figure}[tb]
    \centering
    \includegraphics[width=0.8\linewidth]{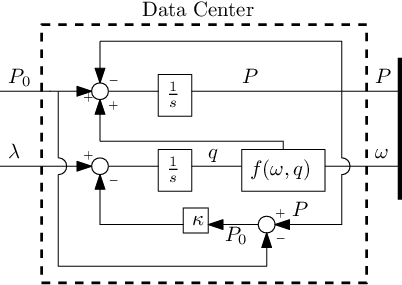}
    \caption{Block diagram of the grid-interactive DC controller.}
    \label{fig:datacenter_control}
    \vspace{-5mm}
\end{figure}

By adopting~\eqref{eq:kP} and~\eqref{eq:control}, we obtain the control architecture illustrated in Figure~\ref{fig:datacenter_control}. The equilibrium of~\eqref{eq:4d_mod} is:
\begin{subequations}
\begin{align}
\bar\omega &= \omega_0, \label{eq:eq_omega}\\
\bar\delta &=
\arcsin\left(
\frac{P_m-P_0-\lambda/\kappa}{\gamma}
\right), \label{eq:eq_delta}\\
\bar P &= P_0+\frac{\lambda}{\kappa}, \label{eq:eq_pdc_val}\\
\bar q &= \frac{\lambda}{\alpha\kappa}. \label{eq:eq_delta_val}
\end{align}
\label{eq:equilibrium}
\end{subequations}
This equilibrium exists provided that
$\left|\frac{P_m-P_0-\lambda/\kappa}{\gamma}\right|\leq 1$.
A few observations from~\eqref{eq:equilibrium} are listed next.

First, the equilibrium DC power demand is the sum of the baseline facility demand and the computational power required to process the incoming workload. In particular, the equilibrium computational power is $\lambda/\kappa$, so the equilibrium total power increases with the workload arrival rate $\lambda$.

Second, the gain $\alpha$ determines the equilibrium computational-work backlog. Increasing $\alpha$ reduces $\bar q$; since backlog is related to request waiting time, $\alpha$ influences latency, although a quantitative latency interpretation requires additional assumptions on the workload and queueing model. More aggressive backlog regulation may also affect closed-loop stability, motivating an explicit analysis of the relationship between the control gains and the region of attraction.

Finally, the gain $\beta$ does not affect the equilibrium, because the frequency deviation vanishes. Nevertheless, $\beta$ influences the transient response to frequency deviations and can therefore affect the stability properties of the coupled system.

It is convenient to introduce the shifted coordinates
$\tilde x=[\tilde\delta,\tilde\omega,\tilde q,\tilde P]^\top$, where
$$
\tilde{\delta} = \delta-\bar{\delta}, \; \;
\tilde{\omega} = \omega-\bar{\omega}, \;\;
\tilde{q} = q-\bar{q}, \;\;
\tilde{P} = P-\bar{P}.$$

Using the equilibrium conditions~\eqref{eq:equilibrium}, and equations~\eqref{eq:kP} and~\eqref{eq:control}, the dynamics can be written as:
\begin{subequations}
\begin{align}
& \dot{\tilde{\delta}} = \tilde{\omega}, \\
& \dot{\tilde{\omega}} = \frac{P_m - \tilde P - \gamma \sin(\tilde{\delta}+\bar{\delta}) - P_0 -\lambda/\kappa}{
J(\tilde{\omega}+\omega_0)
}
-\frac{D}{J}\tilde{\omega},
\\
& \dot{\tilde q} = -\kappa\tilde P, \label{eq:sq}
\\
& \dot{\tilde P}
=
\frac{1}{\tau}
\left[
\beta\tilde{\omega}
+\alpha\tilde q
-\tilde P
\right].
\end{align}
\label{eq:syst_shifted}
\end{subequations}
This coordinate transformation is convenient because it moves the equilibrium to the origin and makes the stability analysis easier. We will use it to  characterize the equilibrium region of attraction in the next section.



\section{Numerical Results}
 
In this section, we simulate the dynamics and study the stability properties of system~\eqref{eq:4d_mod} adopting~\eqref{eq:kP} and~\eqref{eq:control} for different choices of the control parameters $\alpha$ and $\beta$.
All simulations run on an Apple~M4 processor (CPU, 16~GB RAM) and are implemented in Python~3.11.
To run our numerical simulations, we specialize the general DC model introduced in Section~\ref{sec:system_model} to a concrete grid-interactive configuration. We consider a DC with a maximum IT power capacity of $1\,\mathrm{MW}$, whose computational clusters consist of NVIDIA H100 SXM5 GPUs.

According to the NVIDIA H100 SXM5 specifications~\cite{nvidia_h100}, each GPU provides a peak computational throughput of $989.5\,\mathrm{TFLOP/s}$ for dense BF16 Tensor Core operations and has a rated power of $700\,\mathrm{W}$. Based on these specifications, we define the effective computational conversion factor:
\begin{equation}
\kappa
=
\frac{989.5\times10^{12}}
{700}
\approx
1.414\times10^{12}
\frac{\mathrm{FLOP}}{\mathrm{s\,W}}.
\end{equation}

Based on the average utilization estimate reported in~\cite{smith2026united}, we assume a instantaneous IT power consumption equal to $20\%$ of the maximum IT capacity, i.e., $200\,\mathrm{kW}$. This is a simplifying modeling assumption, since GPU utilization does not translate directly into an equivalent fraction of rated power. The corresponding computational workload arrival rate is then:
\begin{equation}
\lambda
=
\kappa \times 200\,\mathrm{kW}
\approx 282.8\,\mathrm{PFLOP/s}.
\end{equation}

We further assume a power usage effectiveness (PUE) of $1.20$, defined as the ratio of total facility power to IT equipment power. At the nominal operating point, this corresponds to $40\,\mathrm{kW}$ of non-IT power consumption, which we model as a constant overhead $P_0=40\,\mathrm{kW}$. It follows from~\eqref{eq:eq_pdc_val} that the total facility power at the equilibrium is:
\begin{equation}
\bar P = 200\,\mathrm{kW}+40\,\mathrm{kW}
=240\,\mathrm{kW}.
\end{equation}

Throughout the results, $P$ is expressed in p.u.\ on the 240~kW equilibrium base ($\bar P = 1$~p.u.), and $q$ in PFLOP. The assumed IT power corresponds to approximately $286$ H100 GPUs operating at their rated power, with the computational throughput estimated using the peak dense BF16 specification.
The remaining model parameters are $J=0.2/\omega_0$, $D=0.16/\omega_0$, $\gamma=2$, $P_m=1$~p.u.\
(so that $\bar\delta=0$), and $\tau=0.5$~s, with control gains $\alpha\in\{0.01,0.1\}$ and $\beta\in\{0,2\}$.
\begin{figure*}[t]
\centering
\includegraphics[width=.32\textwidth]{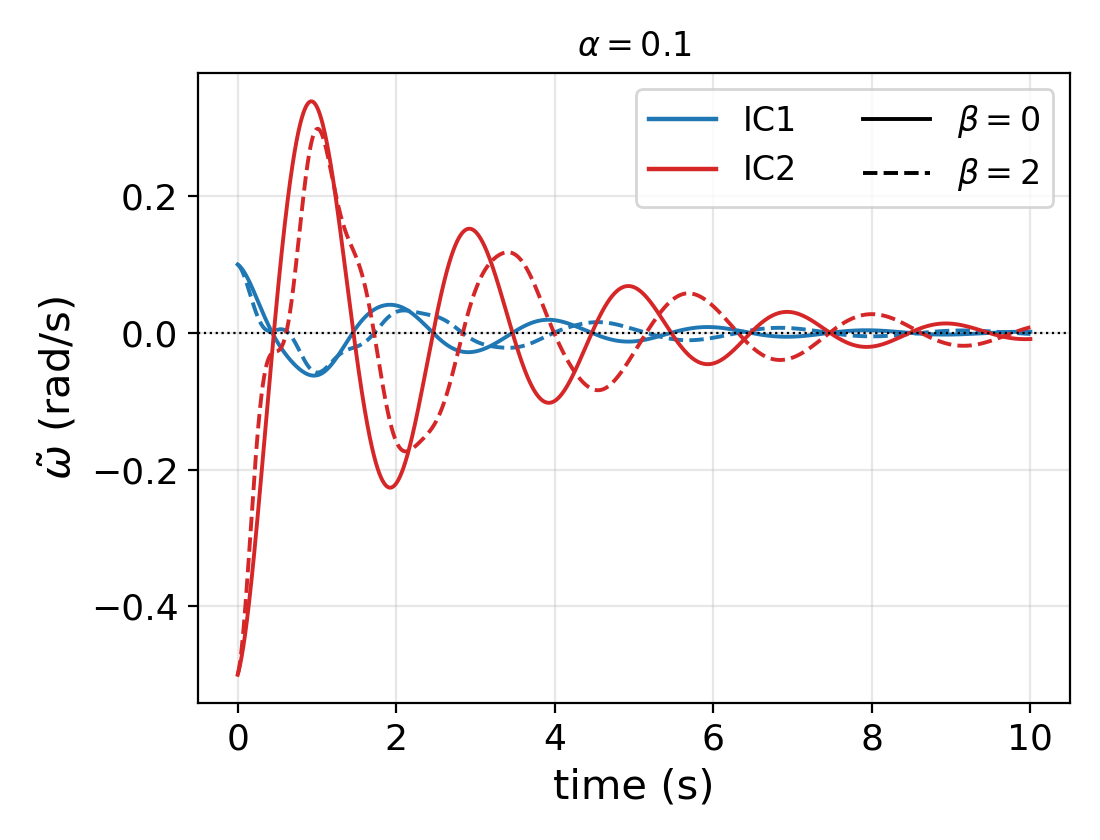}\hfill
\includegraphics[width=.32\textwidth]{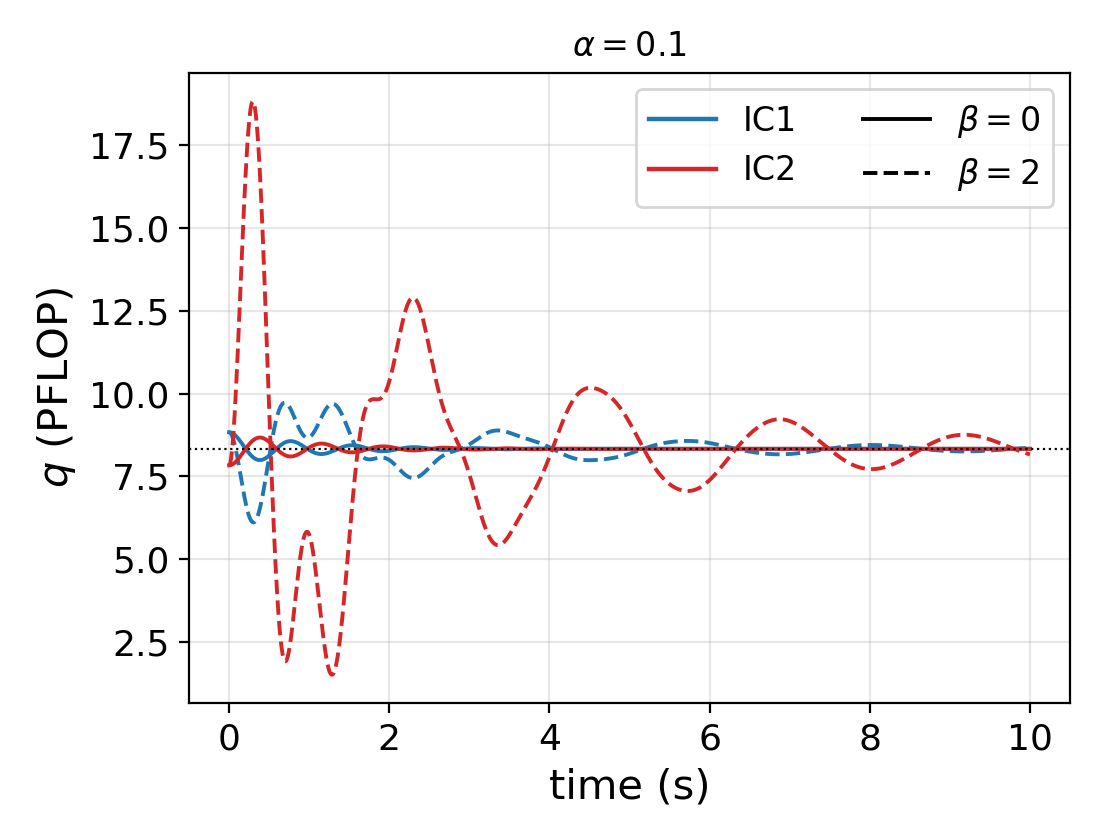}\hfill
\includegraphics[width=.32\textwidth]{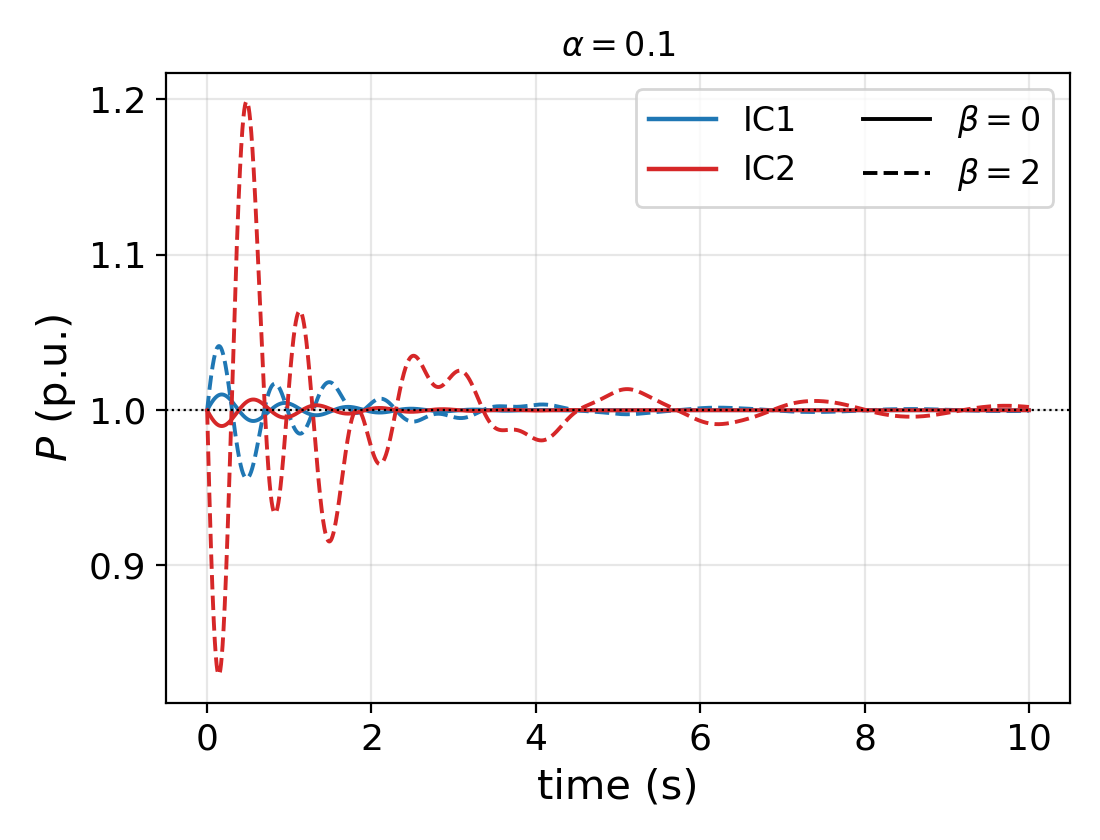}\\[2pt]
\includegraphics[width=.32\textwidth]{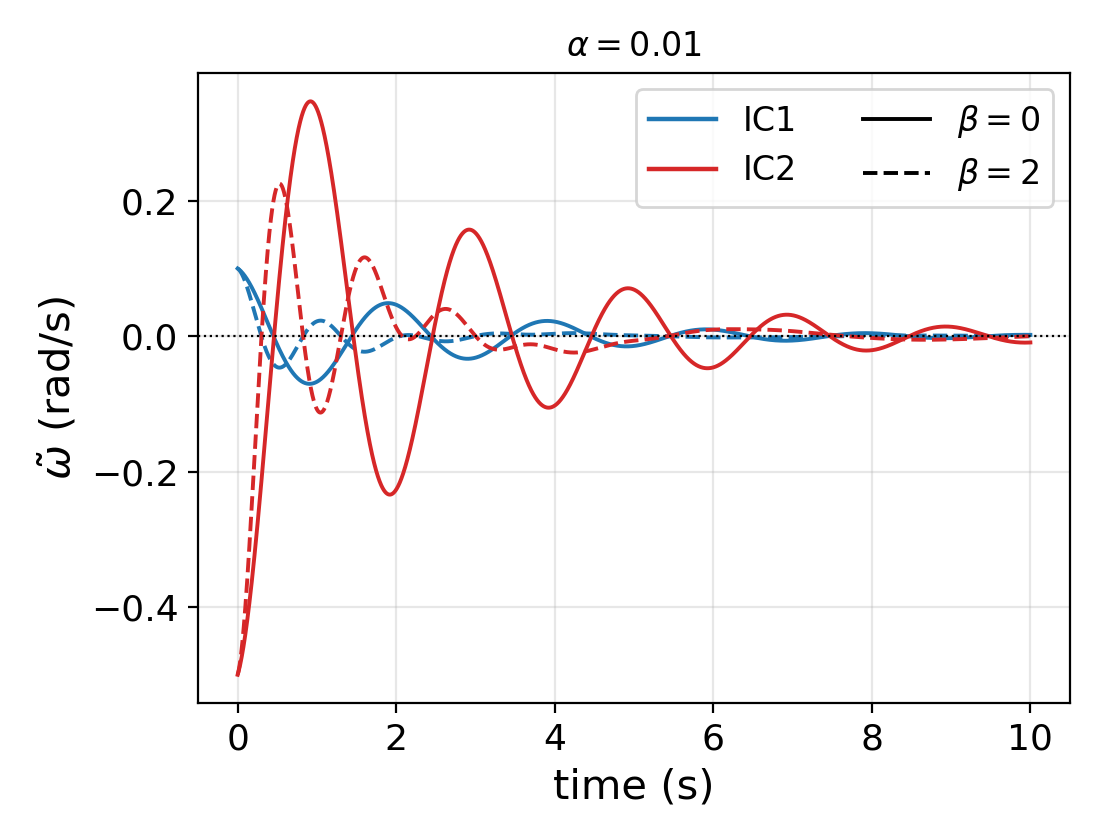}\hfill
\includegraphics[width=.32\textwidth]{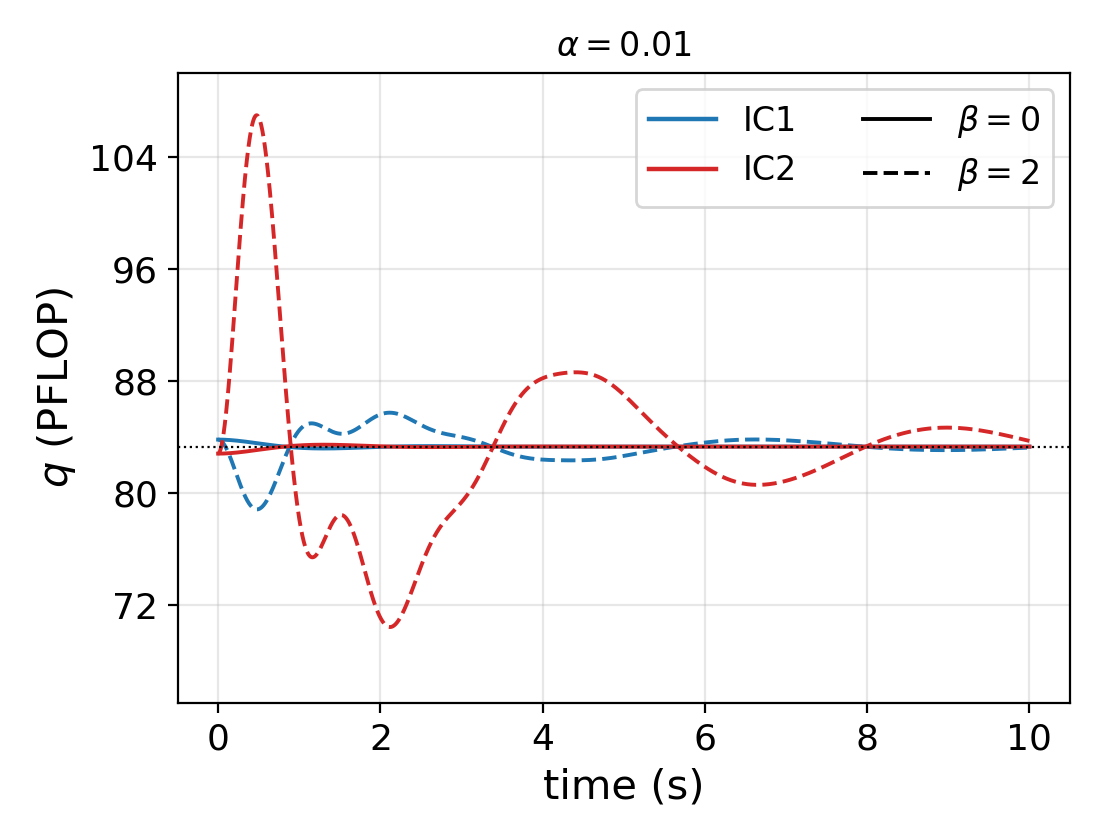}\hfill
\includegraphics[width=.32\textwidth]{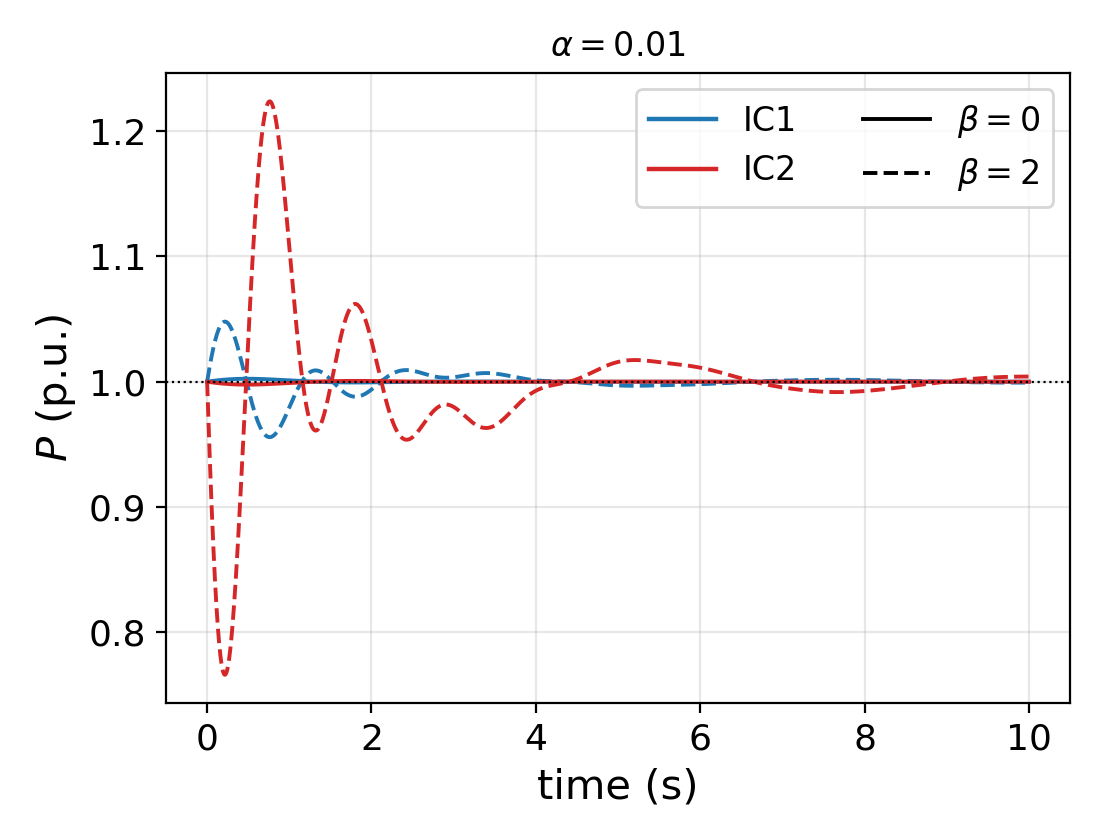}
\caption{Frequency deviation $\tilde\omega$ (left), backlog $q$ (center), and DC power $P$ (right)
for $\alpha=0.1$ (top) and $\alpha=0.01$ (bottom), without droop ($\beta=0$, solid) and with droop
($\beta=2$, dashed), from two initial conditions (IC1, IC2).}
\label{fig:traj}
\end{figure*}

\subsubsection{Time-Domain Simulations}

The model~\eqref{eq:4d_mod} can be used to simulate the dynamic interaction between the DC campus and the power system. We simulate the system trajectories for two different initial conditions, denoted by IC1 and IC2. IC1 represents a scenario in which the system experiences a moderate frequency rise, which could result, for example, from a rapid increase in renewable power generation. IC2, in contrast, represents a severe frequency drop, such as one that may follow a sudden generation loss. For each initial condition, we consider four grid-interactive computing policies defined by the control law~\eqref{eq:control}, corresponding to the gain combinations $\alpha \in \{0.1,0.01\}$ and $\beta \in \{0,2\}$. The resulting trajectories are reported in Figure~\ref{fig:traj}.

Overall, incorporating frequency-responsive control ($\beta=2$) helps damp both the magnitude and duration of frequency deviations, as shown in the left panels of Figure~\ref{fig:traj}. The corresponding DC power trajectories are shown in the right panels. As intended, the DC dynamically adjusts its computational power demand to support grid frequency: demand decreases when the frequency falls below its nominal value and increases when the frequency rises above it. These power adjustments are transient, with the demand returning toward its equilibrium value as the frequency deviation diminishes. Thus, frequency-responsive computing can improve the frequency response of the power system, at the cost of temporarily changing the computational power available to process incoming workloads and, consequently, affecting the computational-work backlog.

The gain $\alpha$ influences both the equilibrium computational backlog and the transient system response. According to~\eqref{eq:eq_delta_val}, the equilibrium backlog is inversely proportional to $\alpha$. In particular, $\alpha=0.01$ yields an equilibrium backlog ten times larger than that obtained with $\alpha=0.1$. However, a larger value of $\alpha$ does not necessarily improve the overall system response. In the simulated cases, the more aggressive backlog-dependent control ($\alpha=0.1$) can make the system more difficult to stabilize, as illustrated by the frequency trajectories in the left panels of Figure~\ref{fig:traj}. This observation highlights the trade-off between reducing the equilibrium computational backlog and preserving favorable closed-loop dynamic behavior.

These simulations suggest that grid-aware computing policies can contribute to frequency support while maintaining a manageable impact on computational performance, provided that the control gains are selected appropriately. In particular, the results illustrate the importance of jointly considering power-system dynamics and computational-work dynamics when designing control policies for grid-interactive DCs.

\subsubsection{Region of Attraction Analysis}

The model~\eqref{eq:4d_mod} also provides a framework for analyzing the stability of the coupled DC--power system under different computing policies. In this section, we estimate the region of attraction (ROA) of the equilibrium~\eqref{eq:equilibrium}. The ROA is defined as the set of initial conditions for which the trajectories of the shifted system~\eqref{eq:syst_shifted} converge to the origin.

We estimate the ROA using the method proposed in~\cite{cavraro2026lyapunov}, adapted to continuous-time systems. The method iteratively expands an initial ROA estimate by constructing a union of sublevel sets associated with learned Lyapunov functions parameterized by neural networks. Unlike conventional approaches that seek to characterize the ROA using a single Lyapunov function, this method combines multiple local Lyapunov-like functions to represent a potentially more complex stability region and obtain less conservative ROA estimates.

\begin{figure}[t]
\centering
\includegraphics[width=0.75\columnwidth]{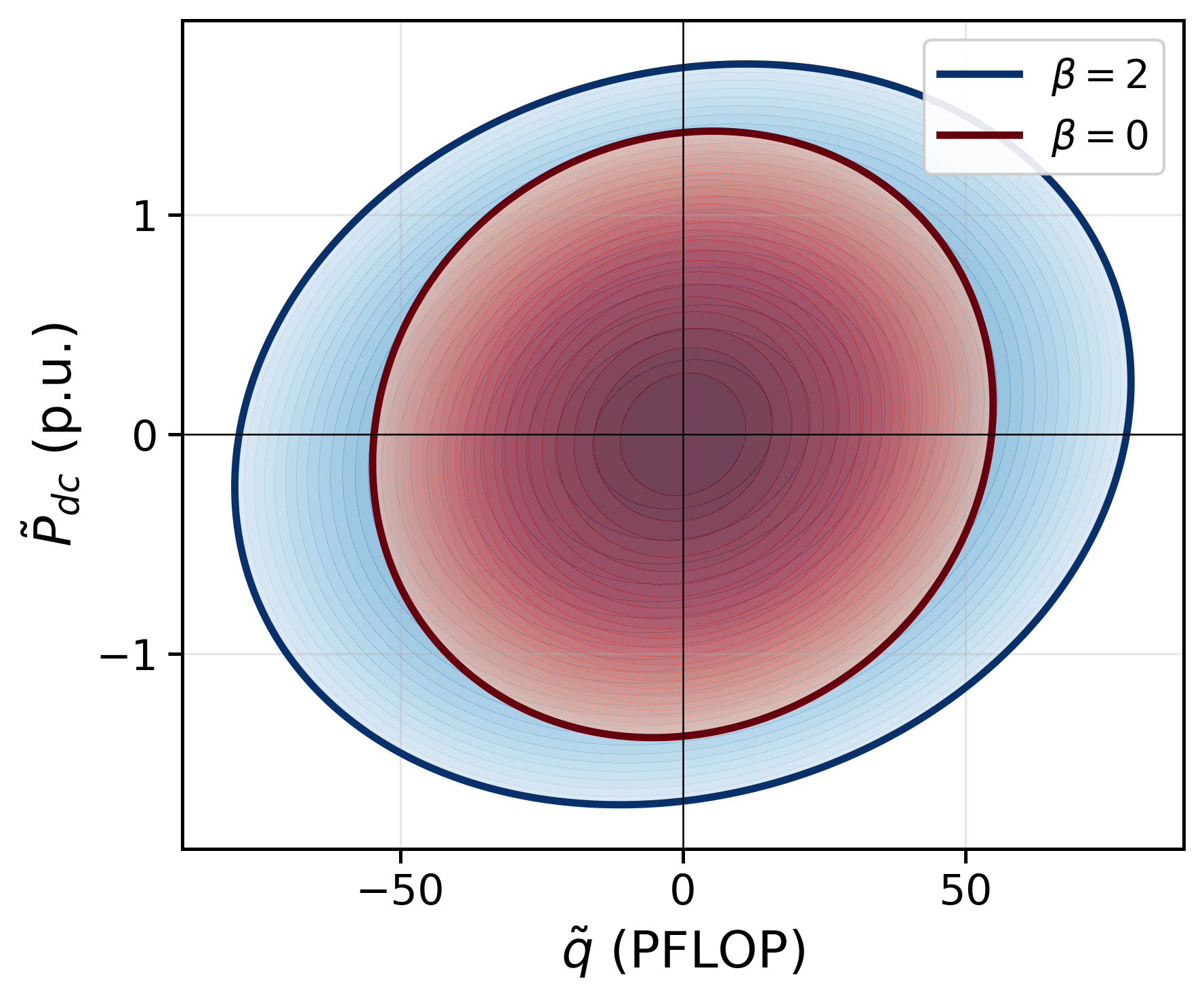}
\caption{ROA in the $(\tilde q,\tilde P_{dc})$ plane for $\beta=0$ (red) and $\beta=2$ (blue), shown as a slice through the equilibrium with $\tilde\delta=\tilde\omega=0$. 
}
\label{fig:roa_Pq}
\end{figure}

Consistent with the time-domain simulations, incorporating the frequency-responsive term in the control law~\eqref{eq:control} improves the estimated stability region. For $\alpha=0.1$, the estimated ROA with $\beta=2$ is approximately $150\%$ larger than that obtained with $\beta=0$. Figure~\ref{fig:roa_Pq} illustrates a two-dimensional slice of the four-dimensional ROA estimates. The comparison shows that introducing frequency-responsive computing can substantially enlarge the estimated set of initial conditions from which the system converges to its equilibrium. These results further demonstrate that the benefits of grid-interactive computing extend beyond transient frequency support to the stability properties of the coupled system.

\section{Conclusion}

This paper presented a four-dimensional dynamical model describing the interaction between a DC campus and a transmission grid, extending the classical SMIB model to account for computational dynamics. The system captures the coupling between electrical variables, including voltage angle, frequency, and DC power demand, and the computational-work backlog, thereby representing how DC operations affect power-system dynamics and how grid conditions influence computational activity.
We showed how grid-aware computing policies can modulate computational power demand in response to grid conditions, providing services such as frequency support while accounting for the impact on computational backlog. The proposed model supports both time-domain simulations and stability analysis, including the estimation of the ROA of the system equilibrium. These capabilities enable the systematic assessment of the trade-offs between grid support and computational performance, and provide a foundation for analyzing and designing control policies to facilitate the reliable integration of DCs into power systems.

\bibliographystyle{IEEEtran}
\bibliography{BibliographyGM}

\end{document}